\documentclass[aps,preprint]{revtex4}
\usepackage{amssymb,epsf}
\usepackage{amsmath}
\usepackage{graphicx}
\begin{document}
\title{Topological Black Holes of Gauss-Bonnet-Yang-Mills Gravity}
\author{M. H. Dehghani$^{1,2}$ \footnote{email address:
mhd@shirazu.ac.ir}, N. Bostani$^3$\footnote{email address:
nbostani@ihep.ac.cn} and R. Pourhasan$^1$}
\affiliation{$^1$Physics Department and Biruni Observatory, College of Sciences, Shiraz University, Shiraz 71454, Iran\\
$^2$ Research Institute for Astrophysics and Astronomy of Maragha (RIAAM), Maragha,
Iran\\$^3$ Key Laboratory of Particle Astrophysics, Institute of High Energy Physics,
Chinese Academy of Sciences, Beijing 100049, China}

\begin{abstract}
We present the asymptotically AdS solutions of Gauss-Bonnet gravity
with hyperbolic horizon in the presence
of a non-Abelian Yang-Mills field with the gauge semisimple group
$So(n(n-1)/2-1,1)$. We investigate the
properties of these solutions and find that the non-negative mass solutions
in 6 and higher dimensions are real everywhere with spacelike singularities. They present
black holes with one horizon and have the same causal structure as the Schwarzschild
spacetime. The solutions in 5 dimensions
or the solutions in higher dimensions with negative mass are not real everywhere. In these cases, one needs a
transformation to make the solutions real. These solutions
may present a naked singularity, an extreme black hole, a black hole with
two horizons, or a black hole with one horizon.

\end{abstract}

\maketitle
\section{Introduction}

The field equations which describe gravitation in classical interaction with
the Yang-Mills fields are highly complicated nonlinear equations. However,
considering the static spherically symmetric spacetime and specific gauge
group such as $SU(N)$, the field equations will reduce to a system of
ordinary differential equations. In particular, Bartnik and McKinnon \cite
{BarMc} were the first who discovered a class of asymptotically flat spherically
symmetric solitonic solutions of Einstein gravity in the presence of
a non-Abelian  $SU(2)$ Yang-Mills field in four dimensions, numerically.
This attracted a great deal of attentions in the
properties of the Einstein-Yang-Mills (EYM) gauge theories for their rich geometrical and physical
structure. Indeed, it was a surprising discovery since for a long time it
had been conjectured that there are no solitonic solutions in EYM models,
based on the fact that there are no solitons in pure gravity \cite{Heusler},
pure YM theories \cite{DeserYM} nor EYM in three dimensional spacetime \cite
{DeserEYM}.

Historically, Yasskin \cite{Yasskin} found the first black hole solutions of
the EYM equations and he conjectured that these solutions were the only
possible ones. However, about two decades later, it was shown that this
conjecture would be false. In fact, soon after the
discovery of solitons for the $SU(2)$ EYM theory \cite{BarMc}, static
spherically symmetric black hole solutions of this model were found
numerically \cite{Bizon,Kunzle}. It is well-known now that the EYM theory contains
{\it hairy} black hole solutions (see \cite{Volkov} for a detailed review).
Among EYM black hole solutions, those in asymptotically AdS space seem to be
more interesting. This is due to the fact that all the asymptotically flat
solutions of pure EYM  gravity, discovered to date, are unstable \cite{BrStrau}.
This intimates that,
although the no-hair theorem is violated in this case, its spirit is
preserved: stable black hole are still simple objects characterized by a few
parameters. But the situation is different in the case of asymptotically AdS
spacetime for which it is possible to find at least some stable black holes with
hair \cite{BjoHoso, Wins}. The non-linear nature of field equations of the
EYM theory make them so intricate, and most of the work in this context have
been done numerically. Nevertheless, some analytical solutions have been
found by considering some particular ansatz to make the field equations
simpler. One of this ansatz is the familiar Wu-Yang ansatz which was
originally introduced in $N=4$ field theory \cite{WuYang} and was applied by
Yasskin \cite{Yasskin} to find an exact black hole solution. Recently, this
ansatz has been used to find the EYM solutions in higher dimensions \cite
{Halil}.

Accepting a nonlinear theory in the matter side of the gravitational field
equations invites nonlinear terms in the geometrical side of the field
equations too, and therefore we extend the geometry side of the field
equations from Einstein tensor to the so-called Lovelock tensor \cite
{lovelock}. Lovelock gravity is the most general classical theory of
gravity, for which the field equations are generally covariant and contain
at most second order derivatives of the metric. In this paper, we
investigate the topological black holes of second order Lovelock (Gauss-Bonnet) gravity coupled
with the non-Abelian gauge field with $So(n(n-1)/1-1,1)$ gauge symmetry. These kinds of black holes
in Gauss-Bonnet-Yang-Mills (GBYM) gravity with spherical horizon have been introduced
in \cite{Hal2}.

The outline of this paper is as follows. In Sec. \ref{Fiel}, we introduce
the field equations of the $(n+1)$-dimensional Gauss-Bonnet gravity in the presence of YM fields
with a semisimple gauge group. In
Sec. (\ref{5dim}), we obtain the 5-dimensional solution of this theory and
investigate its properties. Section \ref{ndim} is devoted to the higher-dimensional solutions of GBYM
gravity and their properties. We finish our paper with some
concluding remarks.

\section{Field equations\label{Fiel}}

The model which will be discussed here is an ($n+1$)-dimensional
GBYM system for an $N$-parameters gauge group $%
\mathcal{G}$, which is assumed to be at least semisimple with structure
constants $C_{bc}^{a}$. The metric tensor of the gauge group is
\begin{equation}
\Gamma _{ab}=C_{ad}^{c}C_{bc}^{d},  \label{Gam}
\end{equation}
where the Latin indices $a$, $b$.... go from $1$ to $N$, and the repeated
indices is understood to be summed over. According to Cartan's criteria the
determinant of $\Gamma _{ab}$ is not zero, and therefore one may define
\begin{equation}
\gamma _{ab}\equiv -\frac{\Gamma _{ab}}{\left| \det \Gamma _{ab}\right| ^{1/N}%
},  \label{Cmet}
\end{equation}
where $\left| \det \Gamma _{ab}\right| $ is the positive value of
determinant of $\Gamma _{ab}$. The action of $(n+1)$-dimensional GBYM
gravity may be written as
\begin{equation}
I_{\mathrm{GBYM}}=\int d^{n+1}x\sqrt{-g}\;[\mathcal{L}_{1}+\alpha _{2}%
\mathcal{L}_{2}-2\Lambda -\gamma _{ab}F_{\mu \nu }^{(a)}F^{(b)\mu \nu }],
\label{Act1}
\end{equation}
where $\mathcal{L}_{1}=R$ is just the Einstein-Hilbert Lagrangian, $\mathcal{%
L}_{2}=R_{\mu \nu \rho \sigma }R^{\mu \nu \rho \sigma }-4R_{\mu \nu }R^{\mu
\nu }+R^{2}$ is the Gauss-Bonnet Lagrangian, $\alpha _{2}$ is Gauss-Bonnet
coefficient with dimension $(\mathrm{length})^{2}$ which is assumed to be
positive, $\Lambda =-n(n-1)/2l^{2}$ is the cosmological constant and $F_{\mu
\nu }^{(a)}$'s are the gauge fields:
\begin{equation}
F_{\mu \nu }^{\left( a\right) }=\partial _{\mu }A_{\nu }^{\left( a\right)
}-\partial _{\nu }A_{\mu }^{\left( a\right) }+\frac{1}{2e}C_{bc}^{a}A_{\mu
}^{\left( b\right) }A_{\nu }^{\left( c\right) }.  \label{gfield}
\end{equation}
In Eq. (\ref{gfield}) $e$ is a coupling constant and $A_{\mu }^{\left(
a\right) }$'s are the gauge potentials. Variation of the action (\ref{Act1})
with respect to the spacetime metric $g_{\mu \nu }$ and the gauge potential $%
A_{\mu }^{(a)}$ yields the GBYM equations as
\begin{eqnarray}
&& G_{\mu \nu }^{(1)}+\alpha _{2}G_{\mu \nu }^{(2)}+\Lambda g_{\mu \nu }
=8\pi T_{\mu \nu },  \label{FE1} \\
&& \nabla _{\mu }F^{\mu \nu }+\frac{1}{e}C_{bc}^{a}A_{\mu }^{(b)}F^{\left(
c\right) \mu \nu }=0\,,  \label{FE2}
\end{eqnarray}
where $G_{\mu \nu }^{(1)}$ is the Einstein tensor, $G_{\mu \nu }^{(2)}$ is
\begin{equation*}
G_{\mu \nu }^{(2)}=2(R_{\mu \sigma \kappa \tau }R_{\nu }^{\phantom{\nu}%
\sigma \kappa \tau }-2R_{\mu \rho \nu \sigma }R^{\rho \sigma }-2R_{\mu
\sigma }R_{\phantom{\sigma}\nu }^{\sigma }+RR_{\mu \nu })-\frac{1}{2}%
\mathcal{L}_{2}g_{\mu \nu },
\end{equation*}
and the stress-energy tensor of the YM field is
\begin{equation}
T_{\mu \nu }=\frac{1}{4\pi }\gamma _{ab}\left( F_{\mu }^{(a)\lambda }F_{\nu
\lambda }^{(b)}-\frac{1}{4}F^{(a)\lambda \sigma }F_{\lambda \sigma
}^{(b)}g_{\mu \nu }\right) .  \label{EMt}
\end{equation}

\section{5-Dimensional Static Solutions\label{5dim}}

The 5-dimensional solution incorporates a logarithmic term unprecedented in
other dimensions, and therefore we shall treat it in some details. The
5-dimensional, static metric with hyperbolic horizon may be written as
\begin{equation}
ds^{2}=-f(r)\;dt^{2}+\frac{dr^{2}}{f(r)}+r^{2}\left\{ d\theta ^{2}+\sinh
^{2}\theta \left( d\varphi ^{2}+\sin ^{2}\varphi d\psi ^{2}\right) \right\} .
\label{Met2}
\end{equation}
Introducing the coordinates
\begin{eqnarray*}
x_{1} &=&r\sinh \theta \sin \varphi \cos \psi , \\
x_{2} &=&r\sinh \theta \sin \varphi \sin \psi , \\
x_{3} &=&r\sinh \theta \cos \varphi , \\
x_{4} &=&r\cosh \theta ,
\end{eqnarray*}
and using the generalized Wu-Yang Ansatz
\begin{eqnarray}
A^{(1)} &=&\frac{e}{r^{2}}\left( x_{1}dx_{4}-x_{4}dx_{1}\right)   \notag \\
&=&-e\sin \varphi \cos \psi d\theta -e\sinh \theta \cosh \theta (\cos
\varphi \cos \psi d\varphi -\sin \varphi \sin \psi d\psi ),  \notag \\
A^{(2)} &=&\frac{e}{r^{2}}\left( x_{2}dx_{4}-x_{4}dx_{2}\right)   \notag \\
&=&-e\sin \varphi \sin \psi d\theta -e\sinh \theta \cosh \theta \left( \cos
\varphi \sin \psi d\varphi +\sin \varphi \cos \psi d\psi \right) ,  \notag \\
A^{(3)} &=&\frac{e}{r^{2}}\left( x_{3}dx_{4}-x_{4}dx_{3}\right)   \notag \\
&=&-e\cos \varphi d\theta +e\sinh \theta \cosh \theta \sin \varphi d\varphi ,
\notag \\
A^{(4)} &=&\frac{e}{r^{2}}\left( x_{1}dx_{2}-x_{2}dx_{1}\right) =e\sinh
^{2}\theta \sin ^{2}\varphi d\psi   \notag \\
A^{(5)} &=&\frac{e}{r^{2}}\left( x_{1}dx_{3}-x_{3}dx_{1}\right)   \notag \\
&=&-e\sinh ^{2}\theta \left( \cos \psi d\varphi -\sin \varphi \cos \varphi
\sin \psi d\psi \right) ,  \notag \\
A^{(6)} &=&\frac{e}{r^{2}}\left( x_{2}dx_{3}-x_{3}dx_{2}\right)   \notag \\
&=&-e\sinh ^{2}\theta \left( \sin \psi d\varphi +\sin \varphi \cos \varphi
\cos \psi d\psi \right) ,  \label{A5}
\end{eqnarray}
one can show that the gauge fields (\ref{A5}) satisfy the YM equation (\ref
{FE2}). The non-zero structure constants of the gauge group are:
\begin{eqnarray}
C_{24}^{1} &=&C_{35}^{1}=C_{41}^{2}=C_{36}^{2}=C_{51}^{3}=C_{62}^{3}=1,
\notag \\
C_{56}^{4} &=&C_{21}^{4}=C_{64}^{5}=C_{31}^{5}=C_{45}^{6}=C_{32}^{6}=1,
\notag
\end{eqnarray}
which show that the gauge group is isomorphic to $So(3,1)$. Using the
definition of the metric tensor of the gauge group, one obtains:
\begin{equation*}
\gamma _{ab}=diag(-1,-1,-1,1,1,1).
\end{equation*}
The non-zero components of energy-momentum tensor (\ref{EMt}) reduce to
\begin{eqnarray}
&&T_{\phantom{t}{t}}^{t}=T_{\phantom{r}{r}}^{r}=-\frac{3e^{2}}{r^{4}}  \notag
\\
&&T_{\phantom{\theta}{\theta}}^{\theta }=T_{\phantom{\varphi}{\varphi}%
}^{\varphi }=T_{\phantom{\psi}{\psi}}^{\psi }=\frac{e^{2}}{r^{4}}
\end{eqnarray}
Defining $\alpha \equiv (n-2)(n-3)\alpha _{2}$, it is a matter of
calculation to show that the solution of the field equation (\ref{FE1}) may
be written as
\begin{equation}
f(r)=-1+\frac{r^{2}}{2\alpha }\left( 1\mp \sqrt{1-\frac{4\alpha }{l^{2}}+%
\frac{12\alpha m}{r^{4}}+\frac{8\alpha e^{2}\ln (r)}{r^{4}}}\right) ,
\label{Fr5}
\end{equation}
where $m$ is the integration constant related to the mass parameter. As one
can see from Eq. (\ref{Fr5}), the solution has two branches with $-$ and $+$
signs. The acceptable sign for which the solution reduces to the solution of
EYM gravity introduced in \cite{Neda} as $\alpha $ goes to zero is the
solution with minus sign.

It is a matter of calculation to show that the solution is asymptotically
AdS with the effective cosmological constant
\begin{equation}
\Lambda _{\mathrm{eff}}=-n(n-1)\alpha \left( 1-\sqrt{1-\frac{4\alpha }{l^{2}}%
}\right) ^{-1},  \label{leff}
\end{equation}
with $n=4$. One may note that $\alpha $ should be less than $l^{2}/4$. The
metric function $f(r)$ given by Eq. (\ref{Fr5}) is real for $r\geq r_{0}$,
where $r_{0}$ is the real root of
\begin{equation}
(l^{2}-4\alpha )r^{4}+12\alpha ml^{2}+8\alpha e^{2}l^{2}\ln (r)=0,
\label{r0}
\end{equation}
given as
\begin{equation*}
r_{0}=\exp \left\{\frac{6m}{e^{2}} -\frac{1}{4e^{2}}\mathrm{LambertW}\left( \frac{%
(l^{2}-4\alpha )}{2\alpha e^{2}l^{2}}e^{-6m/e^{2}}\right) %
\right\} .
\end{equation*}
Thus, one should restrict the spacetime to the region $r\geq r_{0}$, by
introducing a new radial coordinate $\rho $\ as:
\begin{mathletters}
\begin{equation}
\rho ^{2}=r^{2}-r_{0}^{2}\Rightarrow dr^{2}=\frac{\rho ^{2}}{\rho
^{2}+r_{0}^{2}}d\rho ^{2}  \label{Tr}
\end{equation}
With this new coordinate, the above metric becomes:
\end{mathletters}
\begin{equation}
ds^{2}=-f(\rho )dt^{2}+\frac{\rho ^{2}d\rho ^{2}}{(\rho
^{2}+r_{0}^{2})f(\rho )}+(\rho ^{2}+r_{0}^{2})\left\{ d\theta ^{2}+\sinh
^{2}\theta \left( d\varphi ^{2}+\sin ^{2}\varphi d\psi ^{2}\right) \right\} ,
\label{met2}
\end{equation}
where now one should substitute $r=\sqrt{\rho ^{2}+r_{0}^{2}}$\ in Eq. (\ref
{Fr5}).

One can show that the Kretschmann scalar $R_{\mu \nu \rho \sigma }R^{\mu \nu
\rho \sigma }$ diverges at $\rho =0$ $(r=r_{0})$, and therefore there is a
curvature singularity located at $r=r_{0}$. Seeking possible black hole
solutions, we turn to looking for the existence of horizons. In the case of
having black hole solution, the radius of the black hole is obtained by
solving
\begin{equation}
-1+\frac{r_{+}^{2}}{2\alpha }=\frac{r_{+}^{2}}{2\alpha }\sqrt{1-\frac{%
4\alpha }{l^{2}}+\frac{12\alpha m}{r_{+}^{4}}+\frac{8\alpha e^{2}\ln (r_{+})%
}{r_{+}^{4}}}\geq 0,  \label{rh}
\end{equation}
which shows that $r_{+}^{2}\geq 2\alpha $.

\subsection{Extreme black holes:}

First, we consider the conditions of having extreme black holes, for which
the temperature vanishes. The Hawking temperature may be found by analytic
continuation of the metric as
\begin{equation}
T=\frac{2r_{+}^{4}-l^{2}r_{+}^{2}-e^{2}l^{2}}{2\pi
l^{2}r_{+}(r_{+}^{2}-2\alpha )},  \label{Temp}
\end{equation}
and therefore the horizon radius of the extreme black hole located at the
largest real root of the equation $T=0$ given as
\begin{equation}
r_{\mathrm{ext}}^{2}=\frac{l^{2}}{4}\left( 1+\sqrt{1+\frac{8e^{2}}{l^{2}}}%
\right).  \label{rex}
\end{equation}
One can show that $r_{\mathrm{ext}}^{2}$ given by Eq. (\ref{rex}) is
larger than $2\alpha $ for $\alpha \leq l^{2}/4$, and therefore the solution
given in Eq. (\ref{Fr5}) presents an extreme black hole provided the mass
parameter is chosen to be equal to:
\begin{equation*}
m_{\mathrm{ext}}=\frac{r_{\mathrm{ext}}^{4}-l^{2}r_{\mathrm{ext}%
}^{2}-2l^{2}e^{2}\ln r_{\mathrm{ext}}+\alpha l^{2}}{3l^{2}},
\end{equation*}
where $r_{\mathrm{ext}}$ is given by Eq. (\ref{rex}). Numerical calculations
show that the value of $m_{\mathrm{ext}}$ is negative.

\subsection{Non-extreme black holes:}

\begin{figure}
\centering {\includegraphics[width=7cm]{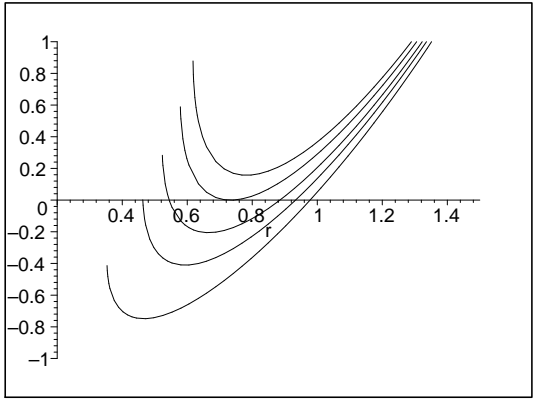}}
\caption{$f(r)$ versus $r$ for $n=4$, $l=1$, $\alpha=.1$, $e=.2$,
$m<m_{\mathrm{ext}}<0$, $m=m_{\mathrm{ext}}<0$, $m_{\mathrm{ext}}<m<m_{\mathrm{c}}$
$m=m_{\mathrm{c}}<0$ and $m>m_{\mathrm{c}}$ from up to down, respectively.} \label{F1}
\end{figure}

Second, we consider the conditions of having non-extreme black holes. In
order to do this, we introduce a critical value for the mass parameter, $m_{%
\mathrm{c}}$, which is the real root of  $f(r_{0}=\sqrt{2\alpha})=0$ given as:
\begin{equation}
m_{\mathrm{c}}=-\frac{1}{3}\left\{\alpha\left(1-\frac{4\alpha}{l^2}\right)+e^{2}\ln (2\alpha)\right\}=0,
\label{r2}
\end{equation}
which is negative. The solution (%
\ref{Fr5}) presents a naked singularity if $m<m_{\mathrm{ext}}$, an extreme
black hole if $m=m_{\mathrm{ext}}$, a black hole with two horizons if $m_{%
\mathrm{ext}}<m\leq m_{\mathrm{c}}$, and a black hole with one horizon
provided $m\geq m_{\mathrm{c}}$. Since $m_{\mathrm{c}}<0$, the solution with
positive mass parameter always presents a black hole with one horizon. Note
that the singularity of the spacetime with $m>m_{\mathrm{c}}$ is spacelike, a
property which does not happen for the 5-dimensional solution of Gauss-Bonnet-Maxwell (GBM)
gravity. Figure \ref{F1} shows the metric function versus the mass parameter
for various values of $m$.

\section{($n+1$)-dimensional static solutions\label{ndim}}

The static metric of an $(n+1)$-dimensional black hole with hyperbolic
horizon may be written as:
\begin{equation}
ds^{2}=-f(r)\;dt^{2}+\frac{dr^{2}}{f(r)}+r^{2}\left( d\theta ^{2}+\sinh
^{2}\theta d\;\Omega _{n-2}^{2}\right)   \label{Metn}
\end{equation}
where $d\Omega _{n-2}^{2}$ is the line element of $(n-2)$-sphere. In order
to obtain the gauge fields, we use the coordinates
\begin{eqnarray*}
x_{1} &=&r\sinh \theta \prod_{j=1}^{n-2}\sin \varphi _{i}, \\
x_{i} &=&r\sinh \theta \cos \varphi _{n-i}\prod_{j=1}^{n-i-1}\sin \varphi
_{j};\text{ \ \ }i=2...n-1, \\
x_{n} &=&r\cosh \theta ,
\end{eqnarray*}
and the ansatz
\begin{eqnarray}
A^{(a)} &=&\frac{e}{r^{2}}\left( x_{i}dx_{n}-x_{n}dx_{i}\right) ;\text{ \ \
\ \ }a=i=1...n-1,  \notag \\
A^{(b)} &=&\frac{e}{r^{2}}\left( x_{i}dx_{j}-x_{j}dx_{i}\right) ;\text{ \ \
\ \ \ \ }i<j,  \label{An}
\end{eqnarray}
where $b$ runs from $n$ to $n(n-1)/2$. It is a matter of calculation to show
that the gauge fields (\ref{An}) satisfy the YM equation, and the the Lie
algebra of the gauge group is $So(n(n-1)/2-1,1)$ with the following $\gamma
_{ab}$:
\begin{equation*}
\gamma _{ab}=\epsilon _{a}\delta _{ab};\text{ \ \ no sum on }a,
\end{equation*}
where $\epsilon _{a}$ is
\begin{equation}
\epsilon _{a}=\left\{
\begin{array}{ll}
-1 & \ \ \ 1\leq a\leq n-1 \\
1 & \ \ \ n\leq (a)\leq \frac{n(n-1)}{2}
\end{array}
\right. .  \label{kappa}
\end{equation}

Using the definition of the energy-momentum tensor (\ref{EMt}), one obtains:
\begin{eqnarray}
&&T_{\phantom{t}{t}}^{t}=T_{\phantom{r}{r}}^{r}=-\frac{(n-2)(n-1)e^{2}}{2r^{4}%
}  \notag \\
&&T_{\phantom{\theta}{\theta}}^{\theta }=T_{\phantom{\phi}{\phi}}^{\phi }=T_{%
\phantom{\psi}{\psi}}^{\psi }=-\frac{(n-2)(n-5)e^{2}}{2r^{4}}.
\end{eqnarray}
Solving the ($n+1$)-dimensional field equation (\ref{FE1}) and using the fact that
the solution should be reduced to the $(n+1)$-dimensional solution of EYM
equation as $\alpha $ goes to zero, one obtains:
\begin{equation}
f(r)=-1+\frac{r^{2}}{2\alpha }\left( 1-\sqrt{1-\frac{4\alpha }{l^{2}}+\frac{%
4(n-1)\alpha m}{r^{n}}+\frac{4(n-2)\alpha e^{2}}{(n-4)r^{4}}}\right) ,
\label{Frn}
\end{equation}
where $m$ is the integration constant related to the mass parameter of the
spacetime. As in the case of 5-dimensional solution, if the following
equation
\begin{equation}
(n-4)(l^{2}-4\alpha )r^{n}+4(n-2)\alpha e^{2}r^{n-4}+4(n-1)\alpha m=0,
\label{r0n}
\end{equation}
has no real solution (non-negative mass), then the metric function (\ref{Frn}%
) is real. For the case that Eq. (\ref{r0n}) has real roots (negative mass), one should
apply the transformation (\ref{Tr}) in order to have a real spacetime, where
$r_{0}$ is now the largest real root of Eq. (\ref{r0n}).

Again, the solution is asymptotically AdS with the cosmological constant
given in Eq. (\ref{leff}). Demanding the absence of conical singularity at
the horizon in the Euclidian sector of the black hole solution, the Hawking
temperature is
\begin{equation}
T=\frac{f^{\prime }(r_{+})}{4\pi }=\frac{%
nr_{+}^{4}-(n-2)l^{2}r_{+}^{2}+(n-4)\alpha l^{2}-(n-2)e^{2}l^{2}}{4\pi
l^{2}r_{+}(r_{+}^{2}-2\alpha )},  \label{tempn}
\end{equation}
which vanishes for the extreme black hole with the horizon radius
\begin{equation}
r_{\mathrm{ext}}^{2}=\frac{(n-2)l^{2}}{2n}\left( 1+\sqrt{1-\frac{%
4n(n-4)\alpha }{(n-2)^{2}l^{2}}+\frac{4ne^{2}}{(n-2)l^{2}}}\right). \label{rexn}
\end{equation}
One can show that $r_{\mathrm{ext}}^{2}$ given by Eq. (\ref{rexn}) is larger
than $2\alpha $ for $\alpha \leq l^{2}/4$, and therefore the solution given
in Eq. (\ref{Frn}) presents an extreme black hole provided the mass parameter
is chosen to be:
\begin{equation}
m_{\mathrm{ext}}=-\frac{(n-2)l^{2}r_{\mathrm{ext}}^{n-4}}{n^{2}}\left( 1-%
\frac{4n\alpha }{(n-2)l^{2}}-\frac{4ne^{2}}{(n-4)l^{2}}+\sqrt{1-\frac{%
4n(n-4)\alpha }{(n-2)^{2}l^{2}}+\frac{4ne^{2}}{(n-2)l^{2}}}\right) ,
\end{equation}
where $r_{\mathrm{ext}}$ is given by Eq. (\ref{rexn}).

\begin{figure}
\centering {\includegraphics[width=7cm]{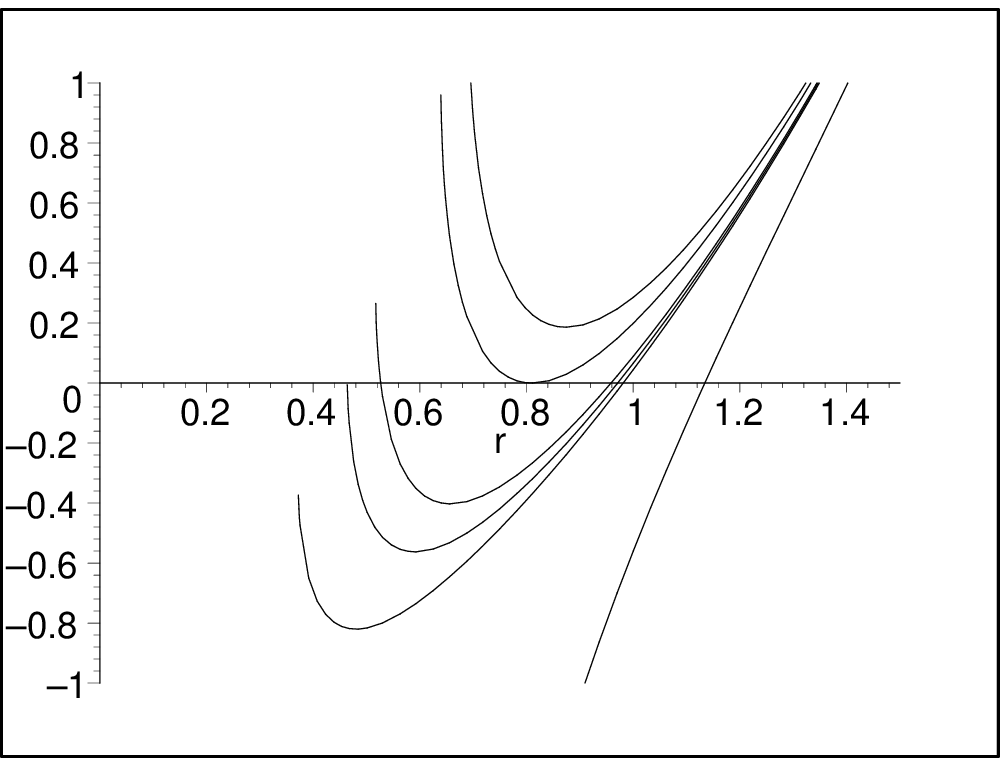} }
\caption{$f(r)$ versus $r$ for $n=6$, $l=1$, $\alpha=.1$, $e=.2$,
$m<m_{\mathrm{ext}}<0$, $m=m_{\mathrm{ext}}<0$, $m_{\mathrm{ext}}<m<m_{\mathrm{c}}$
$m=m_{\mathrm{c}}<0$, $m_{\mathrm{c}}<m<0$ and $m\geq0$ from up to down, respectively.} \label{F2}
\end{figure}

Before discussing the non-extreme black holes, we introduce the critical
mass $m_{\mathrm{c}}$, which is the real root of $f(r_{0}=\sqrt{2\alpha})=0$ give as:
\begin{equation}
m_{\mathrm{c}}=-\frac{(2\alpha)^{(n-4)/2}}{n-1}\left\{\alpha\left(1-\frac{4\alpha}{l^2}\right)+\frac{(n-2)e^2}{n-4}\right\},
\end{equation}
which is negative. Then the
solution given by Eq. (\ref{Frn}) presents a naked singularity if $m<m_{%
\mathrm{ext}}$, an extreme black hole provide $m=m_{\mathrm{ext}}$, a black
hole with two horizon if $m_{\mathrm{ext}}<m<m_{\mathrm{c}}$ and a black
hole with one horizon and spacelike singularity provided $m\geq m_{\mathrm{c}}$. 
Since $m_{\mathrm{c}}<0$, the solution with
positive mass parameter always presents a black hole with one horizon
and spacelike singularity. Figure \ref{F2} shows the metric function versus $r$ for various
values of mass parameter. Also note that $r_{0}$ for non-negative mass is
zero, and the singularity of the solution is spacelike.

\section{Concluding Remarks}

In this paper, we presented the topological solutions of the second order
Lovelock gravity with hyperbolic horizon coupled to a Yang Mills field in the AdS background
and investigated their properties. To obtain the solutions analytically, we used a
spherically symmetric gauge potential with $So(n(n-1)/2-1,1)$ gauge group introduced in Eq. (\ref{An}). Although we
have a matter field, we don't have a new hair. This is due to the special
form of the potential, which causes the matter charge to be zero when one
looks from infinity. We discussed the causal structure of the spacetimes for
different values of the parameters of the solutions such as the Lovelock coefficient, charge and
mass. In 5 dimensions, the metric function is not real everywhere and
therefore we performed a coordinate transformation to make it real in all the
spacetime. However, in higher dimensions the non-negative solution was real
everywhere, and it respects the cosmic censorship hypothesis. That is, the
singularity of non-negative solution is spacelike and it is not naked.
Moreover, the higher-dimensional solutions with negative mass need the
transformation (\ref{Tr}), and may present a naked singularity, an extreme
black hole, a black hole with two horizons or a black hole with one horizon.

\end{document}